\pdfoutput=1

\RequirePackage{fix-cm}

\documentclass[twocolumn]{svjour3}

\smartqed

\usepackage{graphicx}
\usepackage{amssymb,amsmath}
\usepackage{txfonts}
\usepackage{natbib}
\usepackage{hyperref}

\journalname{Astrophysics \& Space Science}

\begin{document}

\title{Comparing the fractal basins of attraction in the Hill problem with oblateness and radiation}

\author{Euaggelos E. Zotos}

\institute{Department of Physics, School of Science, \\
Aristotle University of Thessaloniki, \\
GR-541 24, Thessaloniki, Greece\\
Corresponding author's email: {evzotos@physics.auth.gr}}

\date{Received: 7 June 2017 / Accepted: 7 September 2017}

\titlerunning{Basins of attraction in the Hill problem with oblateness and radiation}

\authorrunning{Euaggelos E. Zotos}

\maketitle

\begin{abstract}

The basins of convergence, associated with the roots (attractors) of a complex equation, are revealed in the Hill problem with oblateness and radiation, using a large variety of numerical methods. Three cases are investigated, regarding the values of the oblateness and radiation. In all cases, a systematic and thorough scan of the complex plane is performed in order to determine the basins of attraction of the several iterative schemes. The correlations between the attracting domains and the corresponding required number of iterations are also illustrated and discussed. Our numerical analysis strongly suggests that the basins of convergence, with the highly fractal basin boundaries, produce extraordinary and beautiful formations on the complex plane.


\end{abstract}

\section{Introduction}
\label{intro}

Over the years a plethora of modifications, regarding the classical restricted three-body problem, have been proposed for modeling more accurately the motion of massless particles in our Solar System \citep[see e.g.,][]{SSR76,SMB85}. By including additional forces to the classical potential we can achieve a more realistic approach for certain applications, such as space flight missions and satellite positioning.

One of the first additions to the potential of the three-body problem was the oblateness coefficient \citep{SSR75}. \citet{OV03} proved that when we take into account the oblateness effects our results, regarding approximations of real satellite orbits in our Solar System, are significantly improved. In the same vein, the restricted three-body problem with oblateness can realistically model several systems in celestial mechanics.

Another interesting case is the restricted three-body problem with effective radiation pressure \citep{S72}. This upgraded model is indeed very important since it takes into consideration the effects of the radiation pressure due to the radiation drag \citep{S80}, thus making the modelling of the system more realistic.

Knowing the basins of convergence is an issue of great importance, since the attracting domains reflect some of the most intrinsic properties of the dynamical system. For obtaining the basins of attraction one should use an iterative scheme and scan a set of initial conditions in order to reveal their final states (attractors). All types of iterative schemes contain the first or even the second order derivatives of the effective potential. The first order derivatives are used in the equations of motion of the test particle, while the second order derivatives enter the variational equations which are mainly used for determining the stability properties of the test particle's dynamics (e.g., for calculating the monodromy matrix of the periodic orbits). On this basis, the iterative schemes combine, in a way, the equations of motion along with the variational equations. In this sense we may claim that the iterative schemes incorporate and combine the dynamics of the test particle's orbit together with the corresponding stability properties and these are exactly the reasons of why we need to know the basins of attraction of a dynamical system.

During the past years a large number of studies has been devoted on the topic regarding the determination of the Newton-Raphson basins of convergence in many types of dynamical systems, such as the Sitnikov problem \citep{DKMP12}, the restricted three-body problem with oblateness and radiation pressure \citep{Z16}, the electromagnetic Copenhagen problem \citep{KGK12,Z17b}, the Copenhagen problem with radiation pressure \citep{K08}, the four-body problem \citep{BP11,KK14,Z17a}, the four-body problem with radiation pressure \citep{APHS16}, the ring problem of $N + 1$ bodies \citep{CK07,GKK09}, or even the restricted 2+2 body problem \citep{CK13}.

In \citet{D10} the Newton-Raphson basins of attraction have been investigated in the Hill problem with oblateness and radiation \citep{MRVK00,MRPD01}. In this work we shall use a large collection of numerical methods in an attempt to reveal and compare the corresponding basins of convergence on the complex plane. Our task is to evaluate the efficiency as well as the convergence speed of the different iterative schemes.

The present article has the following structure: the most important properties of the mathematical model are given in Section \ref{mod}. In Section \ref{numeth} we describe all the iterative schemes and provide details regarding the computational methods we used for the classification. The following Section \ref{bas} contains all the numerical results, regarding the basins of convergence. Our paper ends with Section \ref{disc}, where we emphasize the main conclusions of this work.

\section{Description of the mathematical model}
\label{mod}

For obtaining the Hill problem, with radiating primary (the Sun) and oblate secondary (the planet), we have to make some simple scale changes and also to consider the limiting case where the mass parameter $\mu = m_2/(m_1 + m_2)$ tends to zero $(\mu \rightarrow 0)$ \citep{S67}.

\begin{figure}
\centering
\resizebox{\hsize}{!}{\includegraphics{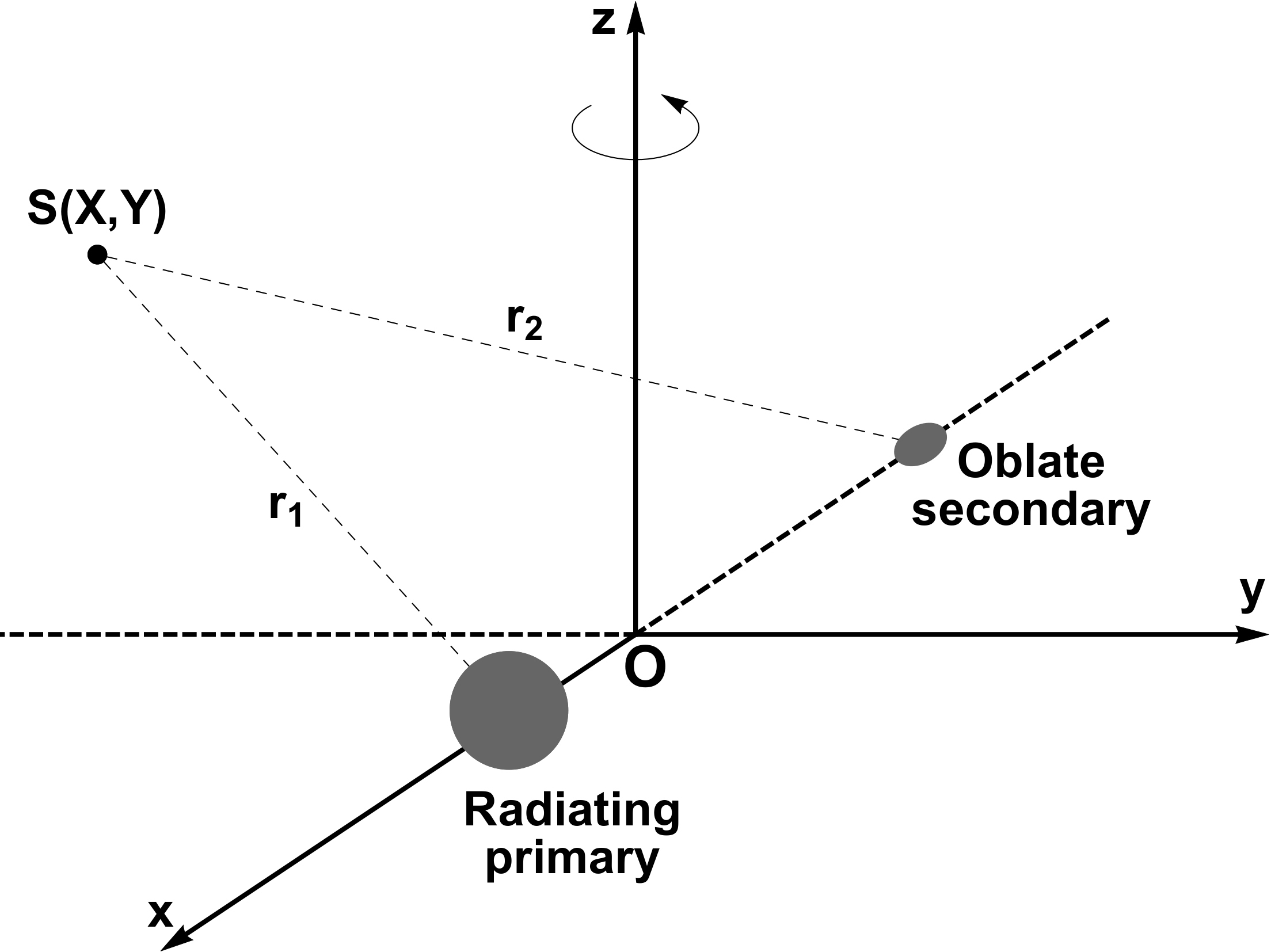}}
\caption{A schematic depicting the space configuration of the circular restricted three-body problem, when the primary body is emitting radiation, while the secondary body is an oblate spheroid.}
\label{scheme}
\end{figure}

According to \citet{SSR75,OV03,AS06} the effective potential function of the circular restricted three-body problem, when the primary is a radiation source and the secondary is an oblate spheroid, whose equatorial plane coincides with the plane of motion (see Fig. \ref{scheme}), is
\begin{equation}
\Omega(X,Y) = \frac{q_1 \left(1 - \mu \right)}{r_1} + \frac{\mu}{r_2} \left(1 + \frac{A_2}{2r_2^2} \right) + \frac{n^2}{2}\left(X^2 + Y^2 \right),
\label{pot}
\end{equation}
where $(X,Y)$ are the coordinates of the test particle, while
\begin{align}
r_1 &= \sqrt{\left(X - \mu\right)^2 + Y^2}, \nonumber\\
r_2 &= \sqrt{\left(X - \mu + 1\right)^2 + Y^2},
\end{align}
are the distances of the test particle from the two bodies and $n = \sqrt{1 + 3A_2/2}$ is the mean motion of the two bodies.

The oblateness coefficient $0 \leq A_2 \ll 1$ is defined as
\begin{equation}
A_2 = \frac{R_E^2 - R_P^2}{5R^2},
\label{obl}
\end{equation}
where $R_E$ and $R_P$ are the equatorial and polar radius of the oblate secondary, respectively, while $R$ is the distance between the two main bodies. Furthermore, the connection between the oblateness $A_2$ and the $J_2$ coefficient is well explained in \citet{A12}.

The radiation factor $q_1 \leq 1$ is defined as
\begin{equation}
q_1 = 1 - \frac{F_R}{F_G},
\label{rad}
\end{equation}
where $F_R$ is the radiation pressure force, while $F_G$ is the gravitational force of the primary.

In a rotating coordinate system the equations of motion are
\begin{align}
\Omega_X &= \frac{\partial \Omega}{\partial X} = \ddot{X} - 2n \dot{Y}, \nonumber\\
\Omega_Y &= \frac{\partial \Omega}{\partial Y} = \ddot{Y} + 2n \dot{X}.
\label{eqmot0}
\end{align}

The next step is to place the origin of the coordinates at the center of the secondary and change the scale of lengths by a factor of $\mu^{1/3}$
\begin{equation}
X = \mu - 1 + \mu^{1/3}x, \ \ \ \ \ Y = \mu^{1/3}y.
\end{equation}

If we apply the above-mentioned transformations to (\ref{pot}), while further scaling the oblateness and radiation coefficients as
\begin{equation}
A_2 = a \mu^{2/3},  \ \ \ \ \ q_1 = 1 - Q \mu^{1/3},
\end{equation}
and take the limit for $\mu \to 0$, we can derive the potential function of the Hill problem with radiating primary and oblate secondary
\begin{equation}
W(x,y) = \frac{3a}{4} + \frac{3x^2}{2} + \frac{1}{r} + \frac{a}{2r^3} - Q x,
\label{pot2}
\end{equation}
with $r = \sqrt{x^2 + y^2}$. More details regarding the derivation of the potential function of the Hill problem with oblateness and radiation are given in \citet{MPD08}.

The corresponding equations of motion read
\begin{align}
W_x &= \frac{\partial W}{\partial x} = \ddot{x} - 2\dot{y} = \left(3 - \frac{1}{r^3} - \frac{3 a}{2r^5}\right) x - Q, \nonumber\\
W_y &= \frac{\partial W}{\partial y} = \ddot{y} + 2\dot{x} = - \left(\frac{1}{r^3} + \frac{3 a}{2r^5} \right) y.
\label{eqmot}
\end{align}

The Hill problem with oblateness and radiation admits the following Jacobi integral of motion
\begin{equation}
J(x,y,\dot{x},\dot{y}) = 2W(x,y) - \left(\dot{x}^2 + \dot{y}^2\right) = \Gamma,
\label{ham}
\end{equation}
where $\Gamma$ is the new Jacobi constant which is related to the classical Jacobi constant $C$ of the restricted three-body problem through the relation
\begin{equation}
C = 3 + \mu^{2/3} \Gamma.
\end{equation}

\section{Numerical methods}
\label{numeth}

The equilibrium points of the system can be found by setting the right-hand sides of Eqs. (\ref{eqmot}) equal to zero
\begin{align}
&3x - \left(1 + \frac{3a}{2r^2}\right)\frac{x}{r^3} - Q = 0, \nonumber\\
&- \left(1 + \frac{3a}{2r^2}\right)\frac{y}{r^3} = 0.
\label{sys}
\end{align}

Following the approach used in \citet{D10} we derive the complex variable form of the right-hand side of the first of Eq. (\ref{sys}), by replacing $x$ with $z$ and $r$ with $\left(\sqrt{z^2} \right)$. Therefore we have
\begin{equation}
f(z;a,Q) = 3z - \left(1 + \frac{3a}{2\left(\sqrt{z^2} \right)^2}\right)\frac{z}{\left(\sqrt{z^2} \right)^3} - Q.
\label{fz}
\end{equation}

\begin{figure}
\centering
\resizebox{\hsize}{!}{\includegraphics{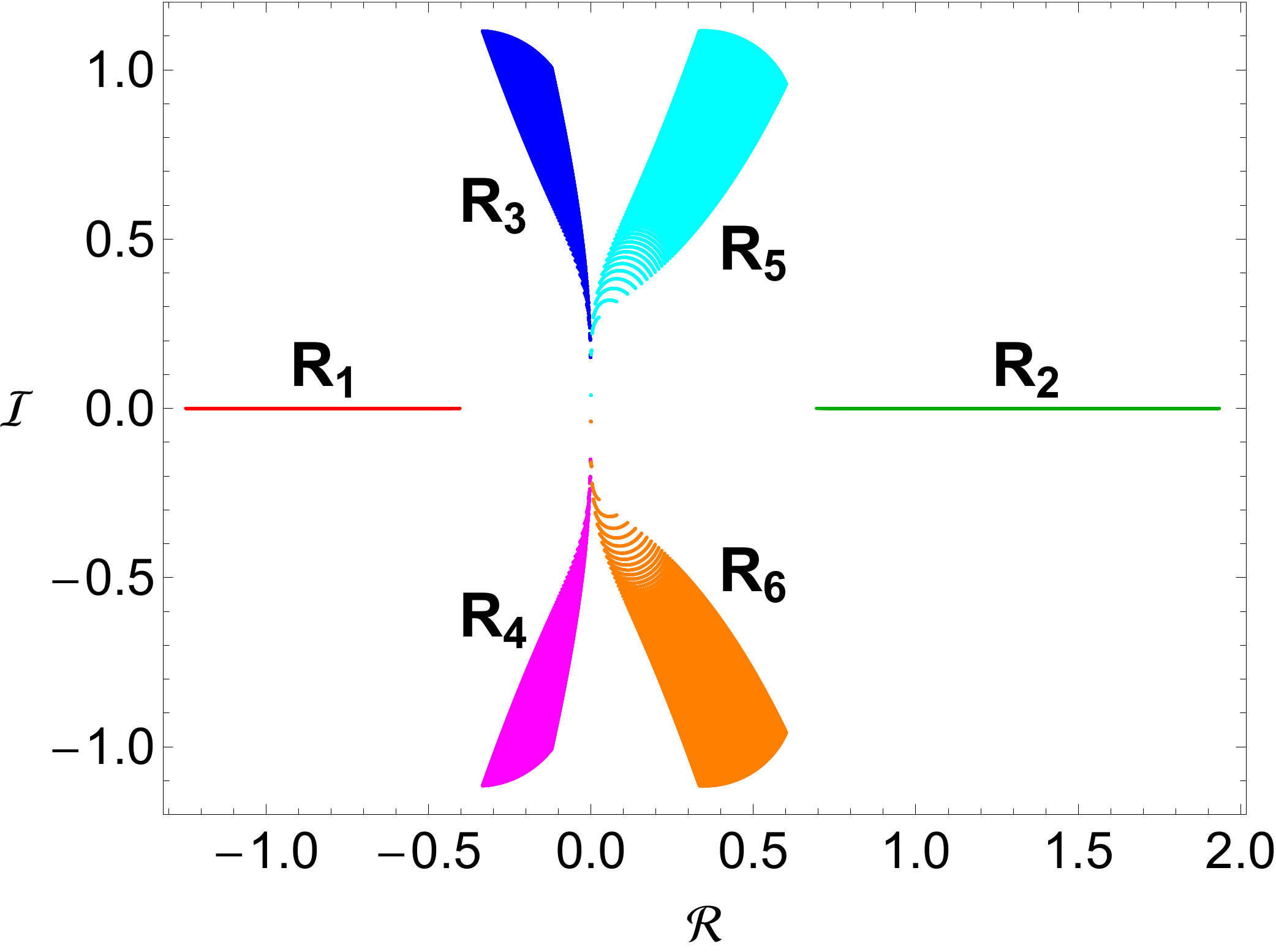}}
\caption{The space evolution of the six roots $R_i$, $i = 1,...,6$ of the complex equation $f(z;a,Q) = 0$, when $a \in [0,5]$ and $Q \in [0,5]$.}
\label{evol}
\end{figure}

Our task is to determine the roots of the complex equation $f(z;a,Q) = 0$. The number as well as the type of the roots strongly depend on the numerical values of the oblateness and the radiation coefficients. In particular, when $a = Q = 0$ and also when $a = 0$ and $Q \neq 0$ there exist two real roots. On the other hands when $a > 0$ we have six roots in total; two real and four complex.

It would be very interesting to monitor the space evolution of the roots as a function of the parameters $a$ and $Q$. In this work the values of both parameters will vary in the interval $[0, 5]$, thus considering only positive values of oblateness and radiation. In Fig. \ref{evol} we present on the complex plane the space evolution of the six roots $R_i$, $i = 1,...,6$, when $a \in [0,5]$ and $Q \in [0,5]$, with $\mathcal{R} = Re[z]$ and $\mathcal{I} = Im[z]$.

Over the years several types of numerical methods have been proposed for solving equations with one variable. In this study we shall consider sixteen methods, with order of convergence $(p)$ varying from 2 to 16. All iterative schemes are explained in the Appendix.

Each numerical algorithm works as follows: The code is activated when we insert a complex number $z = a + ib$, with $\mathcal{R} = a$ and $\mathcal{I} = b$ as initial conditions on the complex plane. The iterative procedure continues until a root (attractor) of the system is reached, with the desired accuracy. If the iterative procedure leads to one of the roots then we say that the method converges for the particular initial condition $(\mathcal{R}, \mathcal{I})$. However, in general terms, not all initial conditions converges to a root of the system. All the initial conditions that lead to a specific final state (attractor) compose the basins of attraction, which are also known as basins of convergence or even as attracting regions/domains. At this point it should be highly noticed that the basins of attraction of the several iterative schemes should not be mistaken, by no means, with the classical basins of attraction which exist in the case of dissipative systems.

Nevertheless, the determination of the basins of attraction of the iterative schemes is a very important task because they reflect, in a way, some of the most intrinsic qualitative properties of the dynamical system. This is true because all the iterative formulae contain the first or even the second order derivative of the function $f(z;a,Q)$.

In order to unveil the basins of convergence we have to perform a double scan of the complex plane. For this purpose we define dense uniform grids of $1024 \times 1024$ $(\mathcal{R},\mathcal{I})$ nodes which shall be used as initial conditions of the numerical algorithms. Of course the initial condition corresponding to the center $(0,0)$ is excluded from all grids, because for this initial condition several terms, entering the iterative formulae, become singular\footnote{Note that the initial condition $(0,0)$ is the only singular point on the complex plane.}. During the classification of the initial conditions we also keep records of the number $N$ of iterations, required for the desired accuracy. Obviously, the better the desired accuracy the higher the required iterations. In our calculations the maximum number of iterations is set to $N_{\rm max} = 500$, while each iterative procedure stops only when an accuracy of $10^{-15}$ is reached.

\section{Basins of attraction}
\label{bas}

In what follows we will try to determine the basins of convergence of the Hill problem, by considering three cases regarding the values of the oblateness and radiation coefficients. For classifying the initial conditions on the complex plane we will use color-coded diagrams (CCDs), where each pixel is assigned a color, according to the final state (root or attractor) of the initial condition. Furthermore, the size of the CCDs (or in other words the minimum and the maximum values of $\mathcal{R}$ and $\mathcal{I}$) is controlled in such a way so as to have, in each case, a complete view, regarding the geometry of the structures produced by the attracting domains.

\subsection{Case I: The classical Hill problem}
\label{ss1}

\begin{figure*}[!t]
\centering
\resizebox{\hsize}{!}{\includegraphics{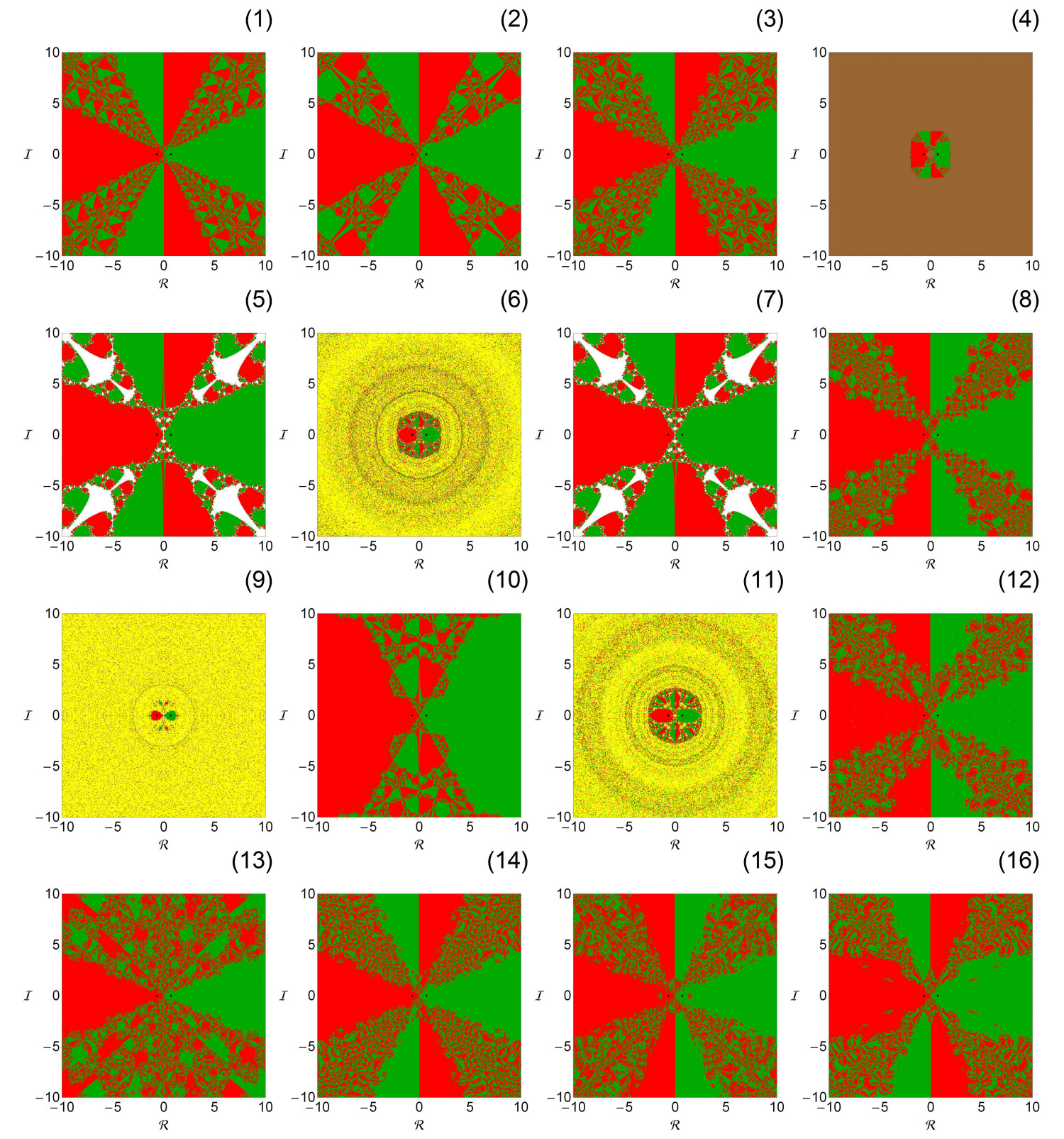}}
\caption{The basins of attraction on the complex plane for the classical Hill problem. The positions of the two real roots are indicated by black dots. The color code is as follows: $R_1$ root (red); $R_2$ root (green); false convergence to zero (brown); false convergence to infinity (yellow); non-converging points (white). Panel identification (numerical method): (1): Newton-Raphson; (2): Halley; (3): Chebyshev; (4): super Halley; (5): modified super Halley; (6): King; (7): Jarratt; (8): Kung-Traub; (9): Maheshwari; (10): Murakami; (11): Neta; (12): Chun-Neta; (13): Neta-Johnson; (14): Neta-Petkovic; (15): Neta $14^{th}$ order; (16): Neta $16^{th}$ order.}
\label{c1}
\end{figure*}

\begin{figure*}[!t]
\centering
\resizebox{\hsize}{!}{\includegraphics{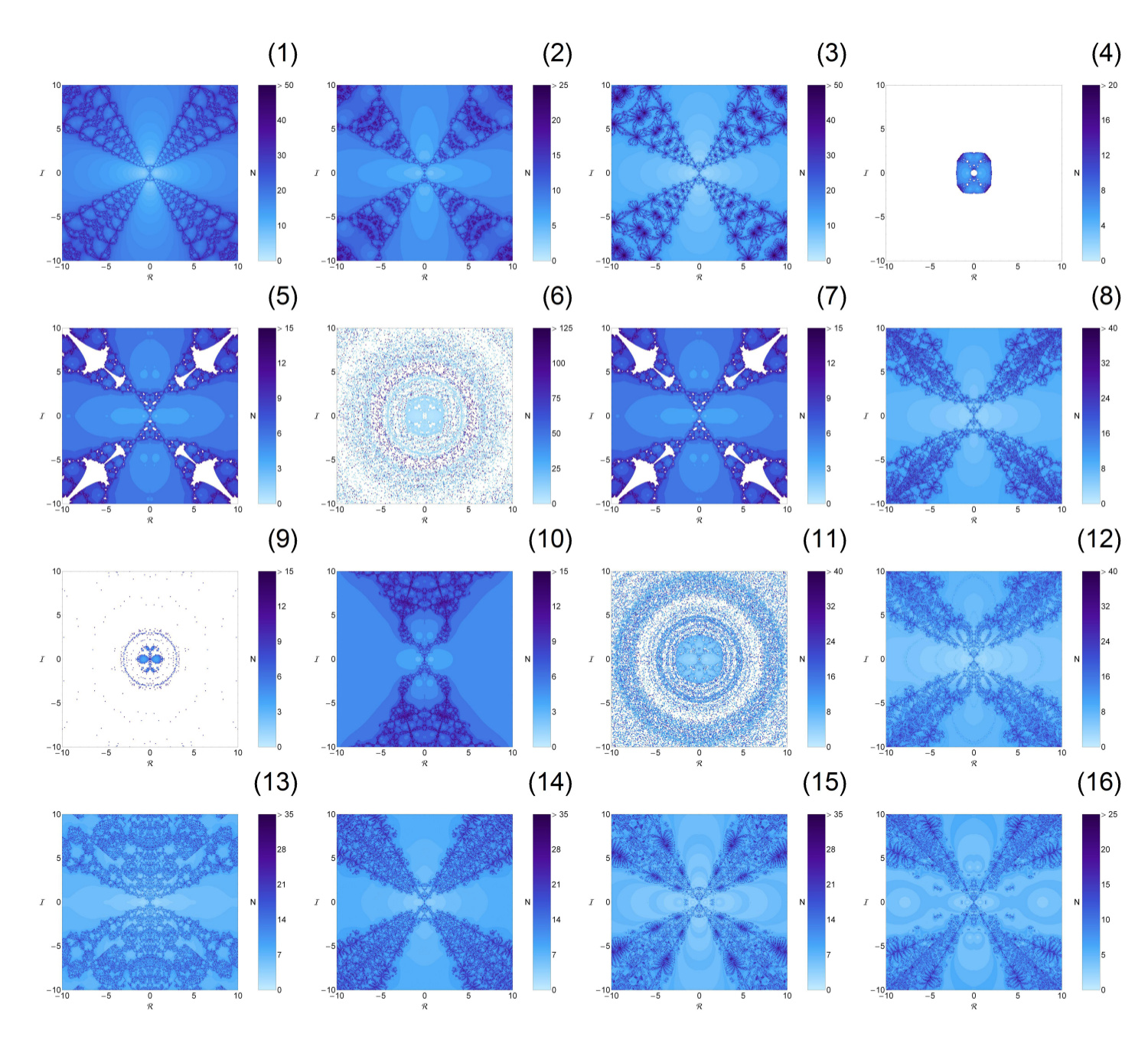}}
\caption{The corresponding distribution of the number $N$ of required iterations for obtaining the basins of attraction shown in Fig. \ref{c1}. False converging and non-converging initial conditions are shown in white. Panel identification (numerical method): (1): Newton-Raphson; (2): Halley; (3): Chebyshev; (4): super Halley; (5): modified super Halley; (6): King; (7): Jarratt; (8): Kung-Traub; (9): Maheshwari; (10): Murakami; (11): Neta; (12): Chun-Neta; (13): Neta-Johnson; (14): Neta-Petkovic; (15): Neta $14^{th}$ order; (16): Neta $16^{th}$ order.}
\label{n1}
\end{figure*}

\begin{figure*}[!t]
\centering
\resizebox{\hsize}{!}{\includegraphics{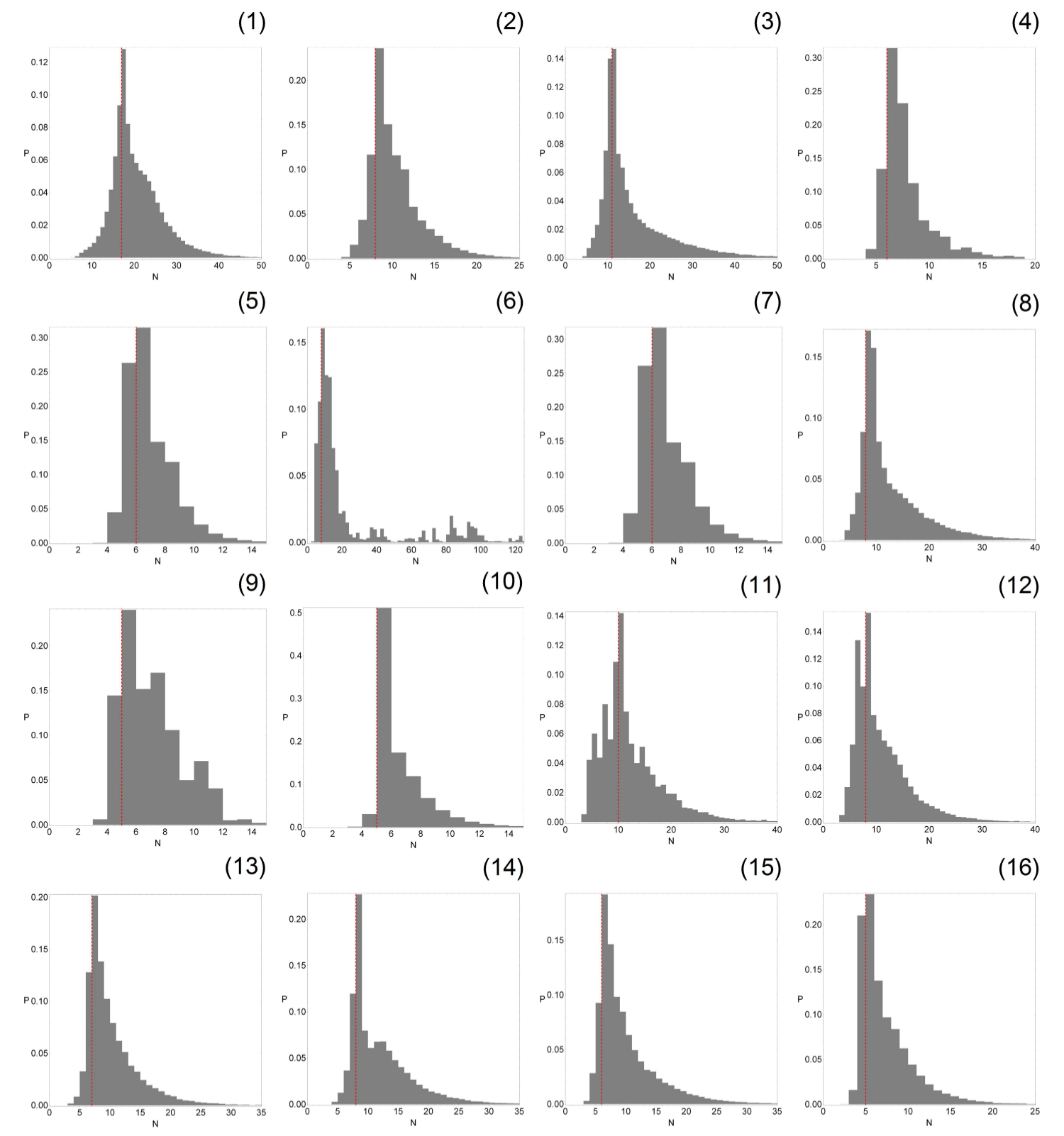}}
\caption{The corresponding probability distribution of the required iterations for obtaining the basins of attraction shown in Fig. \ref{c1}. The vertical dashed red lines indicate, in each case, the most probable number $N^{*}$ of iterations. Panel identification (numerical method): (1): Newton-Raphson; (2): Halley; (3): Chebyshev; (4): super Halley; (5): modified super Halley; (6): King; (7): Jarratt; (8): Kung-Traub; (9): Maheshwari; (10): Murakami; (11): Neta; (12): Chun-Neta; (13): Neta-Johnson; (14): Neta-Petkovic; (15): Neta $14^{th}$ order; (16): Neta $16^{th}$ order.}
\label{p1}
\end{figure*}

\begin{figure*}[!t]
\centering
\resizebox{\hsize}{!}{\includegraphics{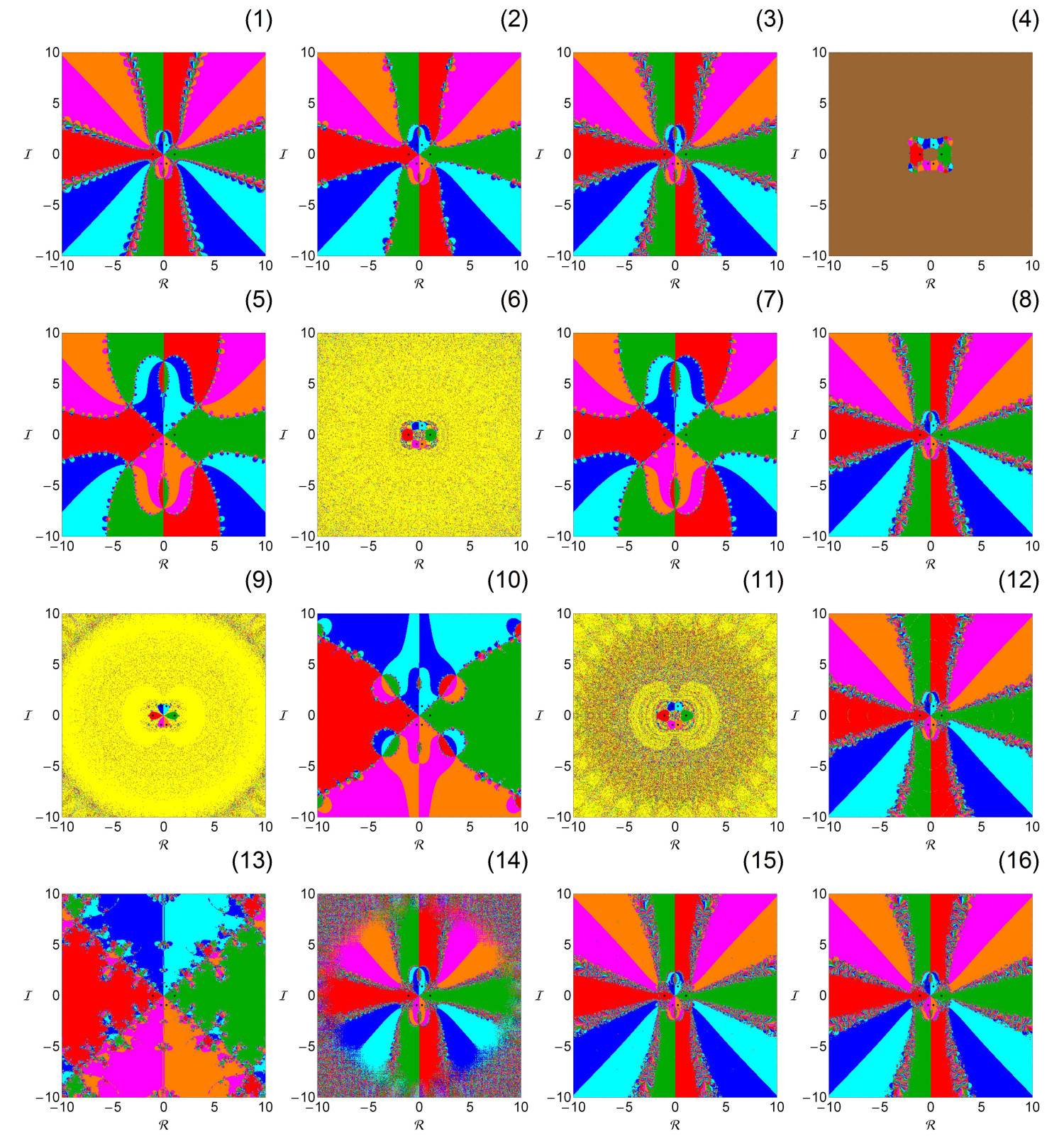}}
\caption{The basins of attraction on the complex plane for the Hill problem with oblateness. The positions of the six roots are indicated by black dots. The color code is as follows: $R_1$ root (red); $R_2$ root (green); $R_3$ root (blue); $R_4$ root (magenta); $R_5$ root (cyan); $R_6$ root (orange); false convergence to zero (brown); false convergence to infinity (yellow); non-converging points (white). Panel identification (numerical method): (1): Newton-Raphson; (2): Halley; (3): Chebyshev; (4): super Halley; (5): modified super Halley; (6): King; (7): Jarratt; (8): Kung-Traub; (9): Maheshwari; (10): Murakami; (11): Neta; (12): Chun-Neta; (13): Neta-Johnson; (14): Neta-Petkovic; (15): Neta $14^{th}$ order; (16): Neta $16^{th}$ order.}
\label{c2}
\end{figure*}

\begin{figure*}[!t]
\centering
\resizebox{\hsize}{!}{\includegraphics{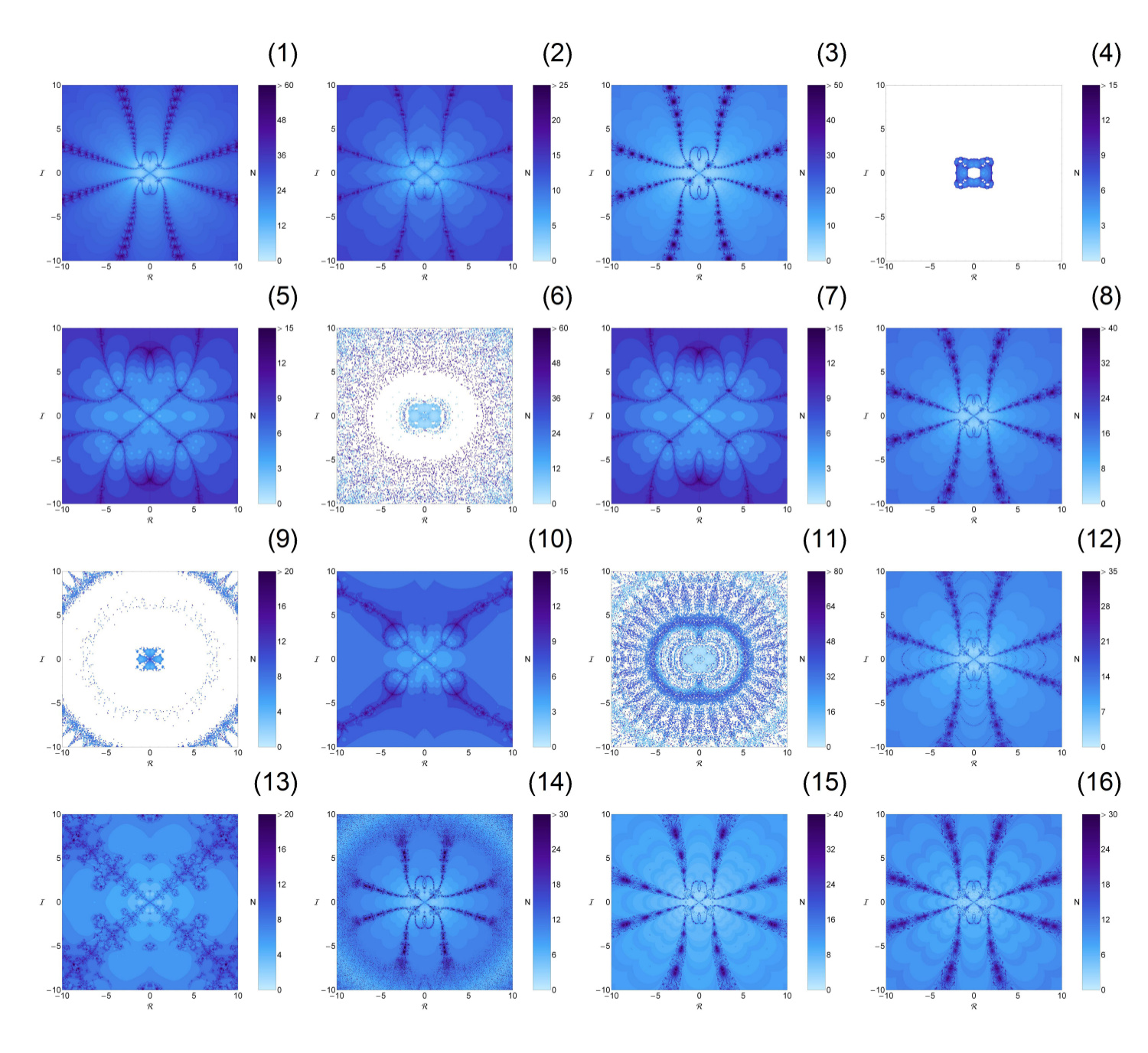}}
\caption{The corresponding distribution of the number $N$ of required iterations for obtaining the basins of attraction shown in Fig. \ref{c2}. False converging and non-converging initial conditions are shown in white. Panel identification (numerical method): (1): Newton-Raphson; (2): Halley; (3): Chebyshev; (4): super Halley; (5): modified super Halley; (6): King; (7): Jarratt; (8): Kung-Traub; (9): Maheshwari; (10): Murakami; (11): Neta; (12): Chun-Neta; (13): Neta-Johnson; (14): Neta-Petkovic; (15): Neta $14^{th}$ order; (16): Neta $16^{th}$ order.}
\label{n2}
\end{figure*}

\begin{figure*}[!t]
\centering
\resizebox{\hsize}{!}{\includegraphics{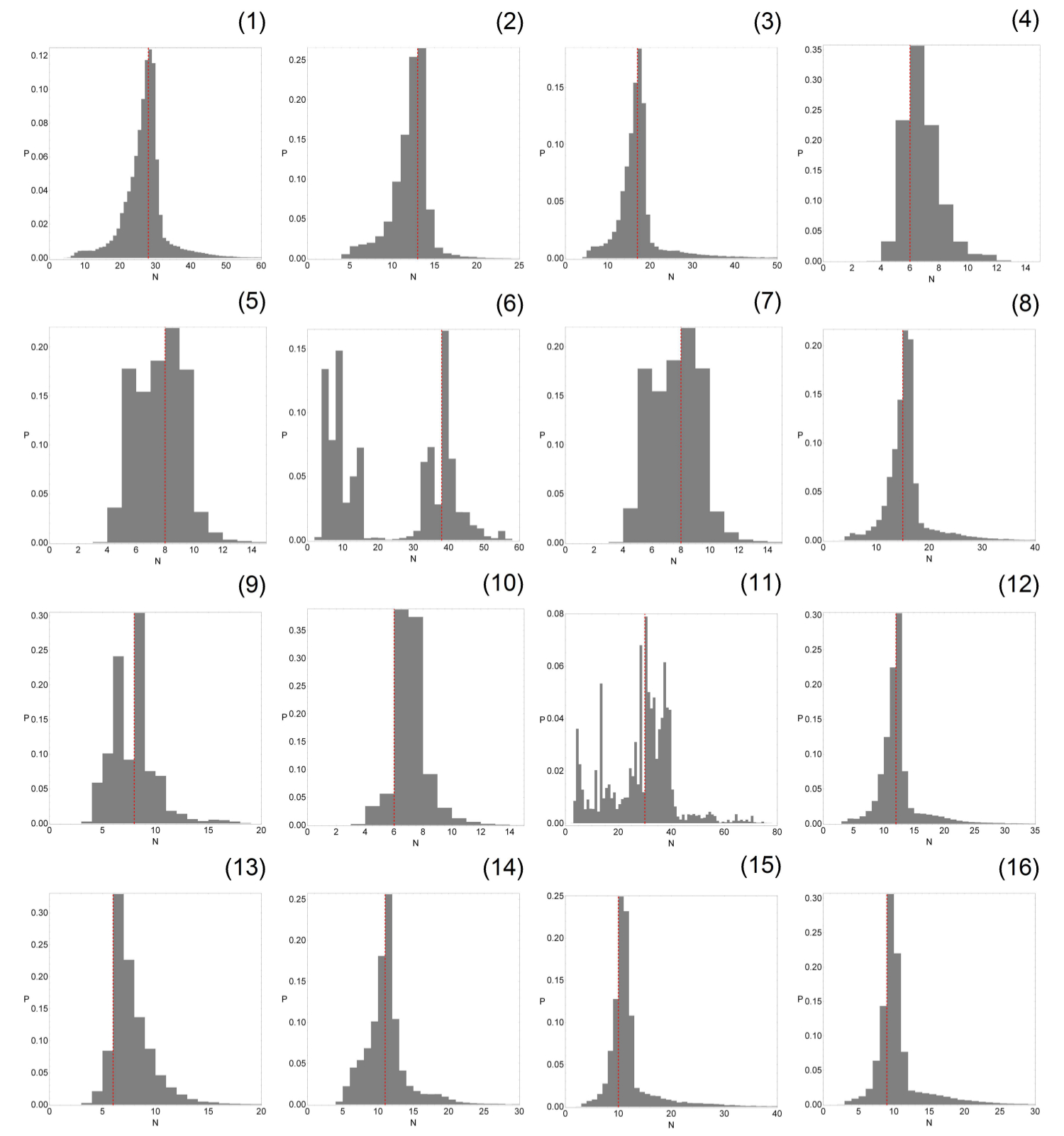}}
\caption{The corresponding probability distribution of the required iterations for obtaining the basins of attraction shown in Fig. \ref{c2}. The vertical dashed red lines indicate, in each case, the most probable number $N^{*}$ of iterations. Panel identification (numerical method): (1): Newton-Raphson; (2): Halley; (3): Chebyshev; (4): super Halley; (5): modified super Halley; (6): King; (7): Jarratt; (8): Kung-Traub; (9): Maheshwari; (10): Murakami; (11): Neta; (12): Chun-Neta; (13): Neta-Johnson; (14): Neta-Petkovic; (15): Neta $14^{th}$ order; (16): Neta $16^{th}$ order.}
\label{p2}
\end{figure*}

\begin{figure*}[!t]
\centering
\resizebox{\hsize}{!}{\includegraphics{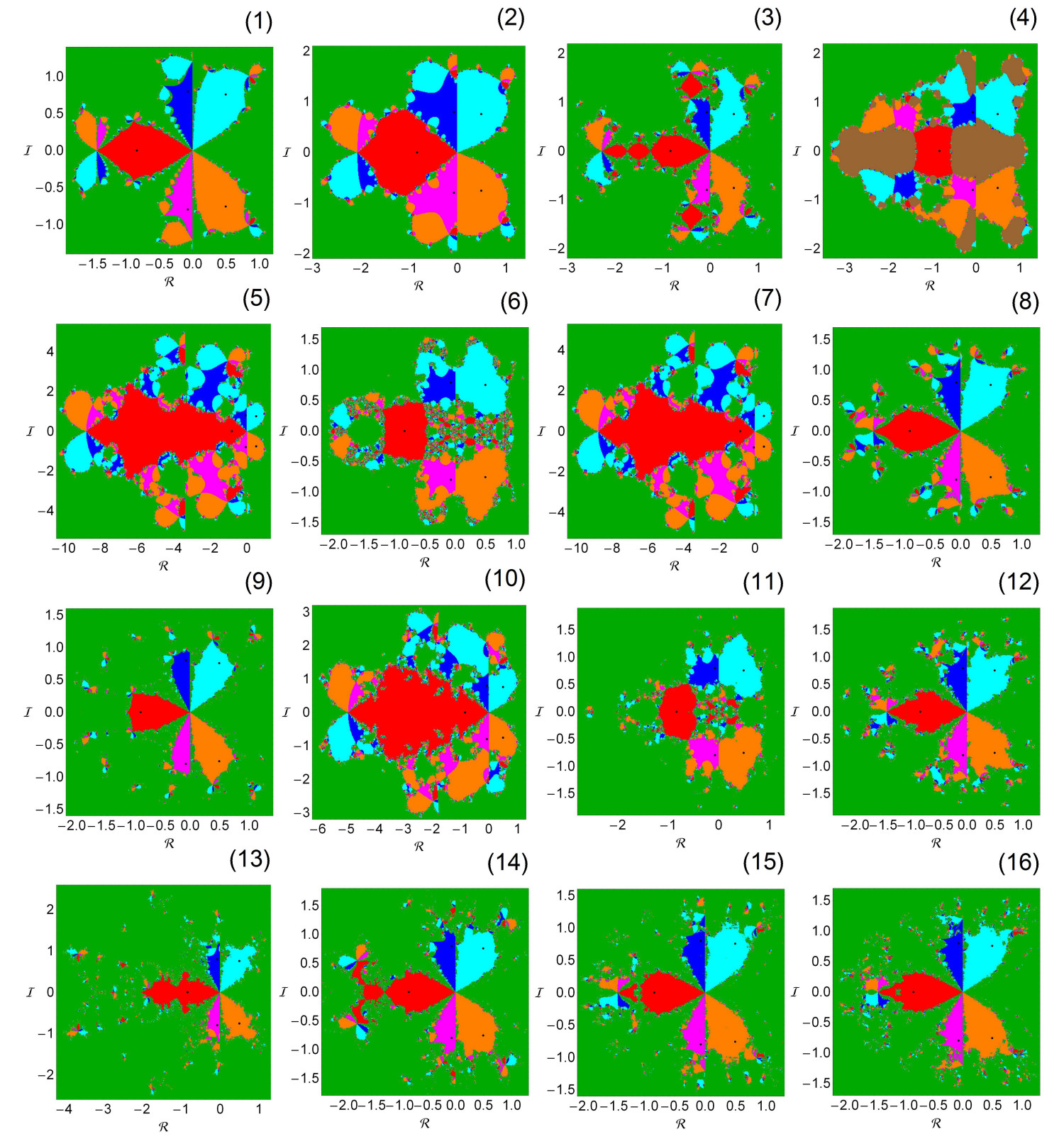}}
\caption{The basins of attraction on the complex plane for the Hill problem with oblateness and radiation. The positions of the six roots are indicated by black dots. The color code is as follows: $R_1$ root (red); $R_2$ root (green); $R_3$ root (blue); $R_4$ root (magenta); $R_5$ root (cyan); $R_6$ root (orange); false convergence to zero (brown); false convergence to infinity (yellow); non-converging points (white). Panel identification (numerical method): (1): Newton-Raphson; (2): Halley; (3): Chebyshev; (4): super Halley; (5): modified super Halley; (6): King; (7): Jarratt; (8): Kung-Traub; (9): Maheshwari; (10): Murakami; (11): Neta; (12): Chun-Neta; (13): Neta-Johnson; (14): Neta-Petkovic; (15): Neta $14^{th}$ order; (16): Neta $16^{th}$ order.}
\label{c3}
\end{figure*}

\begin{figure*}[!t]
\centering
\resizebox{\hsize}{!}{\includegraphics{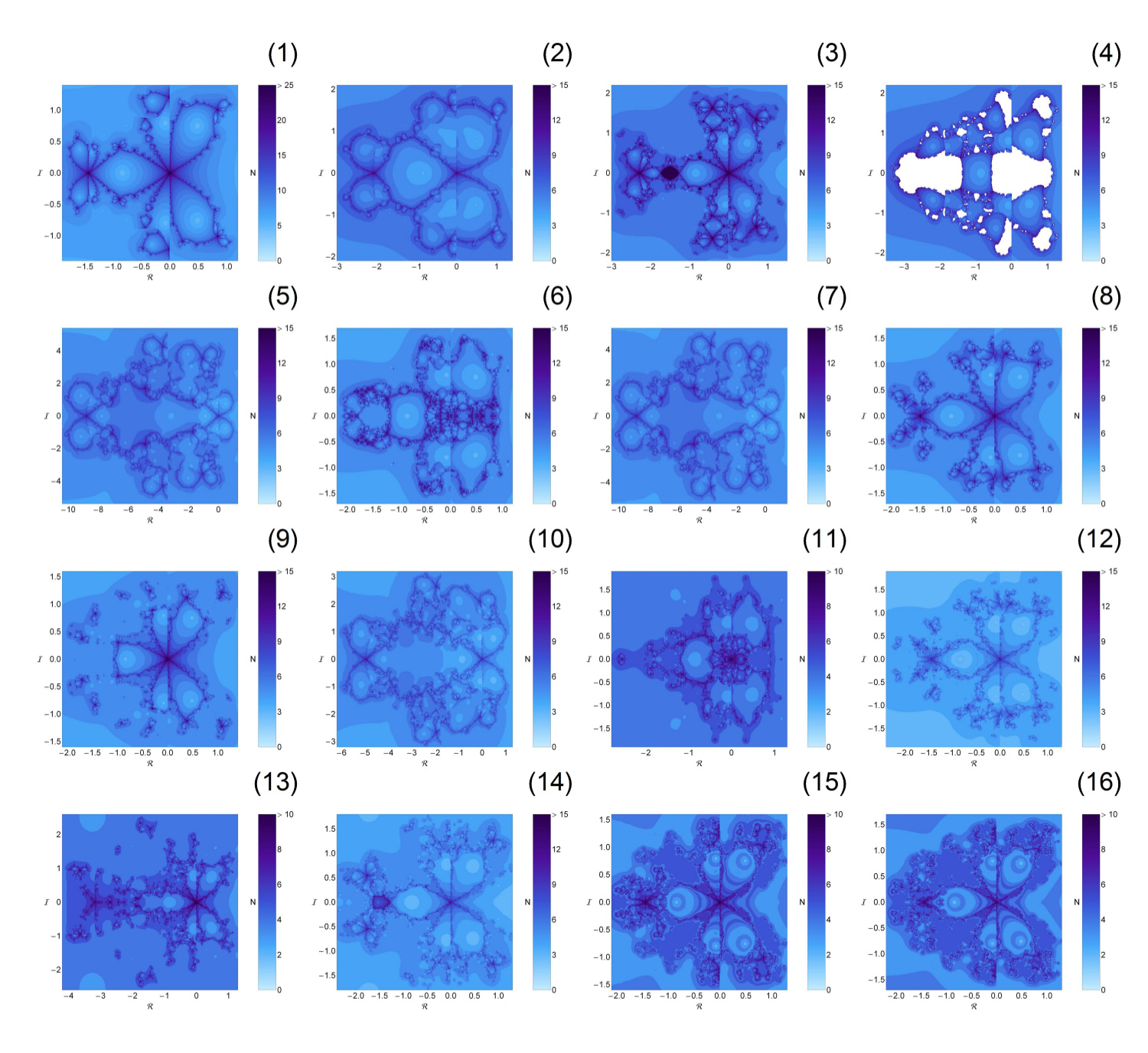}}
\caption{The corresponding distribution of the number $N$ of required iterations for obtaining the basins of attraction shown in Fig. \ref{c3}. False converging and non-converging initial conditions are shown in white. Panel identification (numerical method): (1): Newton-Raphson; (2): Halley; (3): Chebyshev; (4): super Halley; (5): modified super Halley; (6): King; (7): Jarratt; (8): Kung-Traub; (9): Maheshwari; (10): Murakami; (11): Neta; (12): Chun-Neta; (13): Neta-Johnson; (14): Neta-Petkovic; (15): Neta $14^{th}$ order; (16): Neta $16^{th}$ order.}
\label{n3}
\end{figure*}

\begin{figure*}[!t]
\centering
\resizebox{\hsize}{!}{\includegraphics{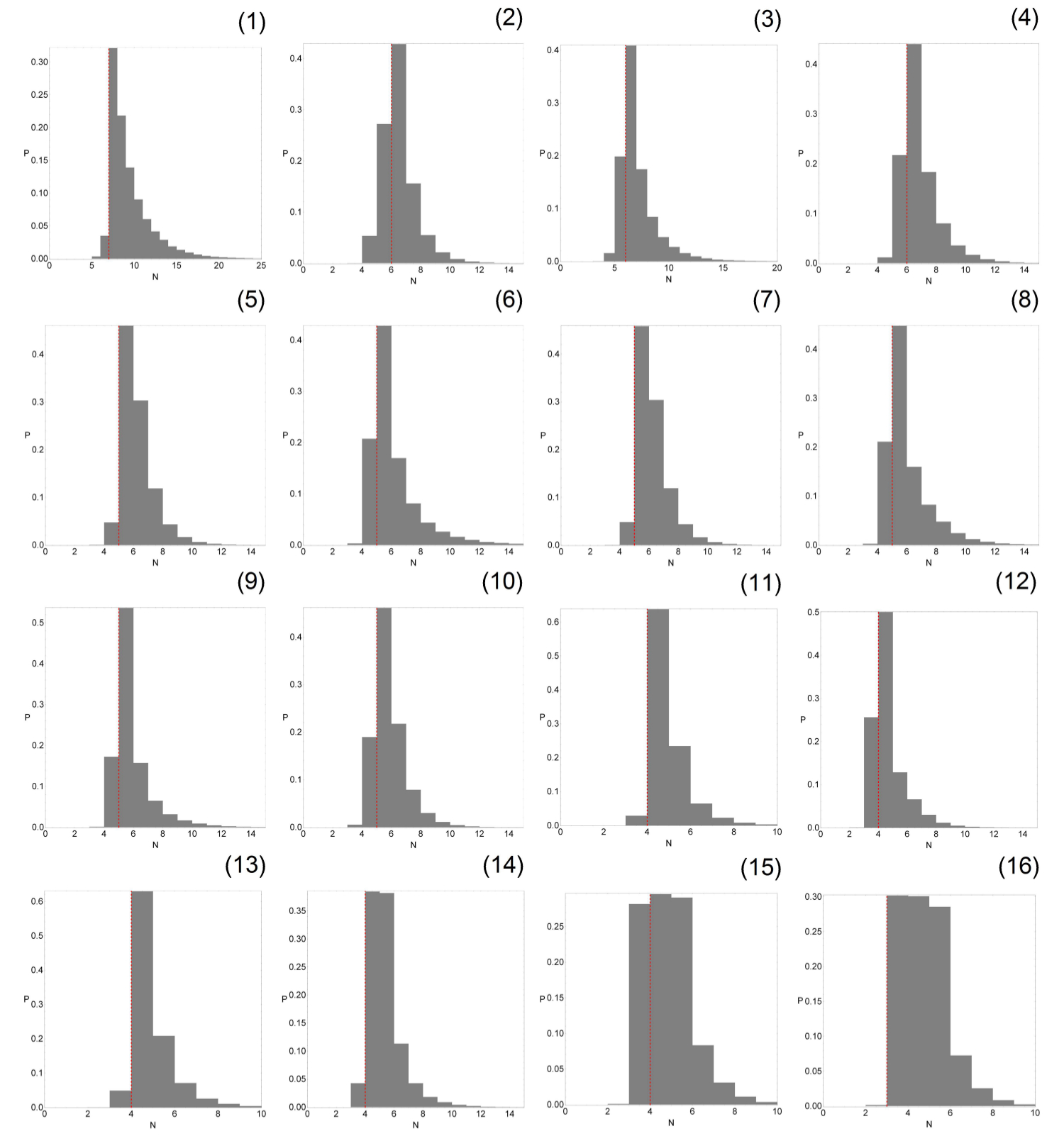}}
\caption{The corresponding probability distribution of the required iterations for obtaining the basins of attraction shown in Fig. \ref{c3}. The vertical dashed red lines indicate, in each case, the most probable number $N^{*}$ of iterations. Panel identification (numerical method): (1): Newton-Raphson; (2): Halley; (3): Chebyshev; (4): super Halley; (5): modified super Halley; (6): King; (7): Jarratt; (8): Kung-Traub; (9): Maheshwari; (10): Murakami; (11): Neta; (12): Chun-Neta; (13): Neta-Johnson; (14): Neta-Petkovic; (15): Neta $14^{th}$ order; (16): Neta $16^{th}$ order.}
\label{p3}
\end{figure*}

The first case under consideration concerns the classical Hill problem, where there is no oblateness or radiation, that is when $a = Q = 0$. In this case Eq. (\ref{fz}) has only two real roots $R_{1,2} = \pm \frac{1}{3^{1/3}}$. For revealing the basins of attraction we scan a square region on the complex plane, where $-10 \leq \mathcal{R}, \mathcal{I} \leq +10$. In Fig. \ref{c1} we present the basins of convergence on the complex plane, using the sixteen iterative schemes explained in the previous section. Our calculations suggest that the numerical methods are divided into two categories: (i) the iterative schemes for which all initial conditions correctly converge to one of the roots and (ii) the iterative schemes for which there is a substantial amount of initial conditions that do not converge to any of the roots. For the classical Hill problem we found that the methods with global convergence are the following: Newton-Raphson, Halley, Chebyshev, Kung-Traub, Murakami, Chen-Neta, Neta-Johnson, Neta-Petkovic, Neta $14^{th}$ order and Neta $16^{th}$ order. For all these methods the basins of attraction corresponding to the two roots extend to infinity, while the patterns on the complex plane are very similar. More precisely, we identify an X-shape pattern which distinguishes between the several well-formed basins of convergence. It is interesting to note that these X-shape patterns are very noisy (highly fractal\footnote{When it is stated that a region is fractal we simply mean that it has a fractal-like geometry, without conducting any additional calculations for computing the fractal dimension as in \citet{AVS01}}) which means the for these initial conditions it is very difficult, or even impossible, to predict their final state (root).

The numerical methods with problematic convergence are the following: super Halley, modified super Halley, King, Jarratt, Maheshwari, and Neta. Our analysis indicate that several types of malfunctions are associated with the corresponding iterative schemes. In particular:
\begin{itemize}
  \item In panel (4) of Fig. \ref{c1} we see that only a small fraction of initial conditions, located near the center, converge to one of the two roots. On the other hand, the vast majority of the surrounded initial conditions display a false convergence to zero $(\mathcal{R} = \mathcal{I} = 0)$.
  \item It is seen in panels (5) and (7) of Fig. \ref{c1} that inside the X-shape regions there are areas composed of initial conditions that do not converge to none of the two roots. Additional numerical computations reveal that for these initial conditions the iterative scheme, after a couple of iterations, is trapped into an infinite loop between the complex numbers $z = \pm 0.01115290 \pm i 0.52780975$. Furthermore, the overall geometry of the complex plane, for both the modified super Halley and Jarratt methods is completely identical.
  \item King's method appears to be highly problematic, according to panel (6) of Fig. \ref{c1}. This is true if we take into account that most of the complex plane is occupied by initial conditions which do not converge to none of the roots. The central region, with the confined basins of convergence, is surrounded by a highly fractal mixture of converging and false converging initial conditions. In this case, apart from initial conditions which converge to zero we have also a substantial amount of initial conditions that converge to extremely large complex numbers, thus indicating convergence to infinity.
  \item There is no doubt that the most pathological iterative scheme is that of the Maheshwari's method. Indeed in panel (9) of Fig. \ref{c1} we observe that only a tiny portion of the complex plane, near the center, corresponds to well-formed basins of attraction. On the other hand, the vast majority of the complex plane is populated by a highly fractal mixture of all types of converging, false converging (to both zero and infinity) and non-converging initial conditions.
  \item Neta's sixth order method is based on King's method (the first two sub-steps are common) and this is exactly why the pattern of complex plane illustrated in panel (11) of Fig. \ref{c1} is very similar to that of panel (6), of the same figure. For both numerical methods false converging initial conditions to zero and infinity dominate the complex plane.
\end{itemize}

The corresponding distribution of the number $N$ of iterations is provided, using tones of blue, in Fig. \ref{n1}. It is observed that initial conditions inside the attracting regions converge relatively fast $(N < 15)$, while the slowest converging points are those in the vicinity of the basin boundaries inside the X-shape patterns. In Fig. \ref{p1} the corresponding probability distribution of iterations is given. The probability $P$ is defined as follows: if $N_0$ initial conditions $(x_0,y_0)$ converge to one of the roots, after $N$ iterations, then $P = N_0/N_t$, where $N_t$ is the total number of initial conditions in every CCD. It was observed that for most of the cases (numerical methods) the most probable number $N^{*}$ of iterations (see the red vertical dashed line in Fig. \ref{p1}) ranges between 5 and 8, while there are also iterative schemes which require more iterations (e.g., the Newton-Raphson, Chebyshev, and Neta methods). Our computations indicate that for almost all numerical methods more than 95\% of the initial conditions converge, to one of the roots, within the first 50 iterations. According to panel (6) of Fig. \ref{p1} only for the King's iterative scheme the corresponding histogram extends to more than 50 iterations.

One should logically expect that as we proceed to iterative schemes of higher order $p$ the most probable number of iterations should be reduced. However it was found that the evolution of $N^{*}$, as a function of $p$, is almost completely random. We believe that it is the strange nature of the specific problem (with all the false converging initial conditions) which is responsible for this abnormal behavior.

\subsection{Case II: Hill problem with oblateness only}
\label{ss2}

This case concerns the Hill problem with oblateness only, that is when $a = 2$ and $Q = 0$. Now the complex equation (\ref{fz}) has two real roots, $R_{1,2}$ and four complex roots $R_{3,4,5,6}$. As in the previous case, a square region on the complex plane, where $-10 \leq \mathcal{R}, \mathcal{I} \leq +10$, is scanned in order to obtain the basins of convergence of the several iterative schemes. Our results presented in the CCDs of Fig. \ref{c2}. We observe that the efficiency of the methods is, in general terms, very similar to that discussed in the previous section. In other words, the problematic iterative schemes, for which there is a substantial amount of false converging initial conditions are the same with the previous case (super Halley, King, Maheshwari, Neta). The main differences with respect to the basins of attraction of the classical Hill problem are the following:
\begin{itemize}
  \item Both the modified super Halley and the Jarratt methods are now completely free of non-converging initial conditions.
  \item In all cases, where the iterative schemes work without malfunctions, all the attracting domains, associated with the six roots, extend to infinity. Furthermore, in all these cases there is a cross-shape noisy pattern which distinguishes between the several well-defined basins of convergence.
  \item In panel (14) of Fig. \ref{c2} we see a very strange pattern, regarding the Neta-Petkovic's method. More precisely, all the basins of convergence form a circular region, while outside of this circular area there is a highly fractal mixture of initial conditions with convergence to all possible roots. Additional computations reveal that this fractal mixture contains also initial conditions with false convergence to zero.
\end{itemize}

In Fig. \ref{n2} we present the corresponding distribution of the number $N$ of iterations, required for obtaining the desired accuracy. Now it becomes more evident that the initial conditions which form the cross-shape patterns are those which need the highest number of iteration in order to converge to one of the roots. The corresponding probability distributions of iterations are given in Fig. \ref{p2}. Once more, the most probable number of iterations $N^{*}$ does not display any regular pattern with respect to the order $p$ of the numerical methods. Looking carefully the panels of Fig. \ref{p2} it is seen that the most smooth histograms are those of the numerical methods for which all the initial conditions converge to one of the roots. On the other hand, the most noisy histograms, with multiple peaks (see e.g., panel (11) of Fig. \ref{p2}), correspond to problematic iterative schemes, with false converging initial conditions.

\subsection{Case III: Hill problem with oblateness and radiation}
\label{ss3}

Our exploration ends with the Hill problem with oblateness and radiation, when $a = 2$ and $Q = 5$. In this case there are again six roots (two real and four complex), regarding the equation (\ref{fz}). The basins of attraction on the complex plane, for the several iterative schemes, are illustrated in Fig. \ref{c3}. We clearly observe that in this case things are very different, with respect to the previous two cases. This is true because in this case the shapes of the attracting domains are only symmetric with respect to the horizontal axis. This implies that the boundaries of the CCDs are no longer fixed for all cases.

For all the numerical methods we found that the basins of convergence corresponding to root $R_2$ (positive real number) extend to infinity, while the attracting domains associated with all the other roots have finite area on the complex plane. Our analysis indicate that in the Hill problem with both oblateness and radiation almost all the iterative schemes work equally well, without abnormalities. Only for the super Halley method we identified a substantial amount of initial conditions $(\mathcal{R}, \mathcal{I})$ for which the iterator displays false convergence to zero. In panel (4) of Fig. \ref{c3} we see that the initial conditions which lead to false convergence are not randomly placed on the complex plane but they form basins of false convergence (brown color).

In Fig. \ref{n3} we can see how the corresponding numbers $N$ of required iteration are distributed on the complex plane, for the numerical methods presented in Fig. \ref{c3}. In panel (3) of Fig. \ref{n3} we observe a very strange behavior, regarding the Chebyshev's method. According to panel (3) of Fig. \ref{c3} the basins of attraction corresponding to root $R_1$ are mainly composed of three lobes, located near the horizontal axis. As we know, the initial conditions which define a basins of convergence are those with the lowest required number of iterations, while the highest number of required iterations correspond to initial conditions located in the vicinity of the basin boundaries. However in panel (3) of Fig. \ref{n3} we observe that all the initial conditions which form the central lobe need substantially much more iterations than the initial conditions of the basin boundaries. So far, we are unable to explain this odd behavior.

The histograms with the corresponding probability distributions of iterations are given in Fig. \ref{p3}. In this case three important phenomena appear, with respect to the previous two cases:
\begin{itemize}
  \item For all the numerical methods the histograms are very smooth, with only one peak.
  \item All numerical methods converge relatively fast, taking into account that more than 95\% of the initial conditions converge within the first 15 iterations (except for the Newton-Raphson method, where the maximum number of required iterations is elevated to 25).
  \item The efficiency of the methods naturally evolves as a function of the order $p$. In other words, the most probable number $N^{*}$ of iterations, which implies the convergence speed of the methods, constantly decreases, as we proceed to iterative schemes of higher order.
\end{itemize}

Before closing this section, we would like to point out that there is also one more case, regarding the values of the parameters of the system. This is the case with radiation and no oblateness, that is when $a = 0$ and $Q \neq 0$. In this case the complex equation (\ref{fz}) has only two real roots (one negative $R_1$ and one positive $R_2$). Our numerical experiments suggest that for all the iterative schemes the basins of attraction corresponding to root $R_1$ are finite, while on the other hand the attracting domains associated with the root $R_2$ extend to infinity. The particular shapes of the basins of convergence are not included here for two main reasons: (i) for saving space and (ii) because the patterns on the complex plane are not so interesting, as in the previous three cases.

\section{Discussion and conclusions}
\label{disc}

The aim of this work was to determine the basins of attraction in the Hill problem with oblateness and radiation, using a large collection of numerical methods. Using the corresponding iterative schemes we managed to reveal the beautiful structures of the basins of convergence on complex plane. The role of the attracting domains is very important since they describe how each initial condition is attracted by the roots, which act as attractors. Our numerical investigation allowed us to monitor the evolution of the geometry of the basins of convergence as a function of the order $p$ of the numerical methods. In addition, the attracting domains have been successfully related with both the corresponding distributions of the number of required iterations and the probability distributions.

As far as we know, there are no previous studies on the basins of attraction of the Hill problem with oblateness and radiation, using numerical methods other than the classical Newton-Raphson scheme. Therefore, all the presented numerical outcomes of the current thorough and systematic analysis are novel and this is exactly the importance and the contribution of our work.

The most important conclusions of our numerical analysis can be summarized as follows:
\begin{enumerate}
  \item In the classical Hill problem, as well as in the case with oblateness only all basins of attraction extend to infinity. In the Hill problem with oblateness and radiation (and also in the case with radiation only) on the other hand, only the basins of convergence, associated to the real root $R_2$, extend to infinity, while all the other attracting domains have finite area on the complex plane.
  \item It was proved that all iterative schemes do not have the same convergence efficiency. In particular we found several types of iterative schemes which display a false convergence to zero or to infinity, for a substantial amount of initial conditions. The most problematic cases were observed when $Q = 0$, while when $Q \neq 0$ only the super Halley method displays the phenomenon of false convergence.
  \item Non-converging initial conditions were identified only in the classical Hill problem and only using the modified super Halley, Jarratt and Maheshwari methods.
  \item The initial conditions with the lowest number of required iterations are those inside the basin of attraction, while those with the highest required number of iterations lie in the vicinity of the basin boundaries. However we came across one specific case where this rule is violated. More precisely, using the Chebyshev's iterative method when $a = 2$ and $Q = 5$ we found a well-formed basin of attraction for which all the corresponding initial conditions require much more iterations, in order to converge, with respect to the required iterations of the initial conditions of the fractal basin boundaries.
  \item For both cases with $Q = 0$ we did not observe any clear increase of the convergence speed of the iterative schemes as a function of the order $p$ of the methods. Only when $Q \neq 0$ it was shown, beyond any doubt, that the most probable number of iterations constantly decreases, as we proceed to higher order methods.
\end{enumerate}

For all the calculation, regarding the determination of the basins of convergence, we used a double precision numerical code, written in standard \verb!FORTRAN 77! \citep{PTVF92}. Furthermore, the latest version 11.1.1 of Mathematica$^{\circledR}$ \citep{W03} was used for creating all the graphical illustration of the paper. For the classification of each set of the initial conditions on the complex plane we needed about 4 minutes of CPU time, using a Quad-Core i7 2.4 GHz PC.

We hope that the present numerical results to be useful in the active field of basins of convergence of numerical methods. Since our present investigation, regarding the attracting domains in the Hill problem with oblateness and radiation, was encouraging it is in our future plans to expand our investigation in other types of dynamical systems.

\section*{Acknowledgments}

The author would like to express his warmest thanks to the anonymous referee for the careful reading of the original manuscript and for all the apt suggestions and comments which allowed us to improve both the quality as well as the clarity of the paper.

\section*{Appendix: Presentation of the iterative schemes}
\label{app}

The sixteen iterative formulae, with order of convergence varying from 2 to 16, are the following:
\begin{itemize}
  \item \textbf{Newton-Raphson's method} $(p = 2)$:
    Newton-Raphson's optimal method \citep[see e.g.,][]{CdB73} is of second order, for simple roots, and the corresponding iterative scheme is given by
    \begin{equation}
    x_{n+1} = x_n - u_n,
    \label{nr2}
    \end{equation}
    where always $u_n = f_n/f'_n$ with $f_n = f(x_n)$ and similarly for the derivatives $f'_n = f'(x_n)$, $f''_n = f''(x_n)$.
  \item \textbf{Halley's method} $(p = 3)$:
    Halley's method \citep{H64} is of third order and the corresponding iterative scheme is given by
    \begin{equation}
    x_{n+1} = x_n - \frac{u_n}{1 - L_f u_n},
    \label{h3}
    \end{equation}
    where $L_f = f''_n/(2f'_n)$.
  \item \textbf{Chebyshev's method} $(p = 3)$:
    Chebyshev's method \citep{T64} is of third order and the corresponding iterative scheme is given by
    \begin{equation}
    x_{n+1} = x_n - \left(1 + \frac{L_f}{2}\right)u_n,
    \label{ch3}
    \end{equation}
    where $L_f = f_n f''_n/(f'_n)^2$.
  \item \textbf{Super Halley's method} $(p = 4)$:
    Super Halley's method \citep{GH01} is of fourth order and the corresponding iterative scheme is given by
    \begin{equation}
    x_{n+1} = x_n - \left(1 + \frac{L_f}{2\left(1 - L_f\right)}\right)u_n,
    \label{sh4}
    \end{equation}
    where $L_f = f_n f''_n/(f'_n)^2$
  \item \textbf{Modified super Halley's method} $(p = 4)$:
    Modified super Halley's optimal method \citep{CH08} is of fourth order and the corresponding iterative scheme is given by
    \begin{align}
    &y_n = x_n - \frac{2}{3}u_n, \nonumber\\
    &x_{n+1} = x_n - \left(1 + \frac{L_f}{2\left(1 - L_f\right)}\right)u_n,
    \label{msh4}
    \end{align}
    where $L_f = \frac{f_n}{(f'_n)^2} \frac{f'(y_n) - f'_n}{y_n - x_n}$.
  \item \textbf{King's method} $(p = 4)$:
    King's method \citep{K73} is of fourth order and the corresponding iterative scheme is given by
    \begin{align}
    &y_n = x_n - u_n, \nonumber\\
    &x_{n+1} = x_n - \frac{\left(f_n\right)^2 + \left(\beta - 1 \right)f_n f(y_n) + \beta \left(f(y_n)\right)^2}{f'_n \left(f_n + \left(\beta - 2\right) f(y_n)\right)},
    \label{k4}
    \end{align}
    where in our experiments we have used $\beta = - 1/2$.
  \item \textbf{Jarratt's method} $(p = 4)$:
    Jarratt's method \citep{J66} is of fourth order and the corresponding iterative scheme is given by
    \begin{align}
    &y_n = x_n - \frac{2}{3}u_n, \nonumber\\
    &x_{n+1} = x_n - \frac{u_n}{2} - \frac{u_n}{2\left(1 + \frac{3}{2}\left(L_f - 1\right)\right)},
    \label{j4}
    \end{align}
    where $L_f = f'(y_n)/f'_n$.
  \item \textbf{Kung-Traub's method} $(p = 4)$:
    Kung-Traub's optimal method \citep{KT74} is of fourth order and the corresponding iterative scheme is given by
    \begin{align}
    &y_n = x_n - u_n, \nonumber\\
    &x_{n+1} = y_n - \frac{f(y_n)}{f'_n \left(1 - L_f \right)^2},
    \label{kt4}
    \end{align}
    where $L_f = f(y_n)/f_n$.
  \item \textbf{Maheshwari's method} $(p = 4)$:
    Maheshwari's optimal method \citep{M09} is of fourth order and the corresponding iterative scheme is given by
    \begin{align}
    &y_n = x_n - u_n, \nonumber\\
    &x_{n+1} = x_n + \frac{1}{f'_n}\left(\frac{f_n^2}{f(y_n) - f_n} - \frac{(f(y_n))^2}{f_n}\right).
    \label{ma4}
    \end{align}
  \item \textbf{Murakami's method} $(p = 5)$:
    Murakami's method \citep{M78} is of fifth order and the corresponding iterative scheme is given by
    \begin{equation}
    x_{n+1} = x_n - a_1 u_n - a_2 w_2(x_n) - a_3 w_3(x_n) - \psi(x_n),
    \label{mu5}
    \end{equation}
    where
    \begin{align}
    &w_2(x_n) = \frac{f_n}{f'(x_n - u_n)}, \nonumber\\
    &w_3(x_n) = \frac{f_n}{f'(x_n + \beta u_n + \gamma w_2(x_n))}, \nonumber\\
    &\psi(x_n) = \frac{f_n}{b_1 f'_n + b_2 f'(x_n - u_n)}.
    \end{align}
    In our experiments we have used the values: $a_1 = 0.3$, $a_2 = -0.5$, $a_3 = 2/3$, $\beta = -1/2$, $b_1 = -15/32$, $b_2 = 75/32$, and $\gamma = 0$.
  \item \textbf{Neta's method} $(p = 6)$:
    Neta's method \citep{N79} is of sixth order and the corresponding iterative scheme is given by
    \begin{align}
    &y_n = x_n - u_n, \nonumber \\
    &z_n = y_n - \frac{f(y_n)}{f'_n} \frac{f_n + \beta f(y_n)}{f_n + \left(\beta - 2\right) f(y_n)}, \nonumber\\
    &x_{n+1} = z_n - \frac{f(z_n)}{f'_n} \frac{f_n - f(y_n)}{f_n - 3f(y_n)}.
    \label{nt6}
    \end{align}
    In \citet{CN12} it was proved that $\beta = -1/2$ is the best choice.
  \item \textbf{Chun-Neta's method} $(p = 6)$:
    Chun-Neta's method \citep{CN12} is of sixth order and the corresponding iterative scheme is given by
    \begin{align}
    &y_n = x_n - u_n, \nonumber\\
    &z_n = y_n - \frac{f(y_n)}{f'_n\left(1 - f(y_n)/f_n \right)^2}, \nonumber\\
    &x_{n+1} = z_n - \frac{f(z_n)}{f'_n\left(1 - f(y_n)/f_n - f(z_n)/f_n \right)^2}.
    \label{cn6}
    \end{align}
  \item \textbf{Neta-Johnson's method} $(p = 8)$:
    Neta-Johnson's method \citep{NJ08} is of eighth order and the corresponding iterative scheme is given by
    \begin{align}
    &y_n = x_n - u_n, \nonumber\\
    &h_n = x_n - \frac{1}{8}u_n - \frac{3f_n}{8f'(y_n)}, \nonumber\\
    &z_n = x_n - \frac{f_n}{f'_n/6 + f'(y_n)/6 + 2f'(h_n)/3}, \nonumber\\
    &x_{n+1} = z_n - \frac{f(z_n)}{f'_n} \frac{f'_n + f'(y_n) + a_2f'(h_n)}{(-1 - a_2)f'_n + (3 + a_2)f'(y_n) + a_2 f'(h_n)},
    \label{nj8}
    \end{align}
    where in our experiments we have used $a_2 = -1$.
  \item \textbf{Neta-Petkovic's method} $(p = 8)$:
    Neta-Petkovic's optimal method \citep{NP10} is of eighth order and the corresponding iterative scheme is given by
    \begin{align}
    &y_n = x_n - u_n, \nonumber\\
    &z_n = x_n - \frac{f(y_n)}{f'_n \left(1 - f(y_n)/f_n \right)^2}, \nonumber\\
    &x_{n+1} = x_n - u_n + c_n f_n^2 - d_n f_n^3,
    \label{np8}
    \end{align}
    where
    \begin{align}
    d_n &= \frac{1}{(f(y_n) - f_n)(f(y_n) - f(z_n))}\left(\frac{y_n - x_n}{f(y_n) - f_n} - \frac{1}{f'_n}\right) \nonumber\\
    &- \frac{1}{(f(y_n) - f(z_n))(f(z_n) - f_n)}\left(\frac{z_n - x_n}{f(z_n) - f_n} - \frac{1}{f'_n}\right), \nonumber\\
    c_n &= \frac{1}{f(y_n) - f_n}\left(\frac{y_n - x_n}{f(y_n) - f_n} - \frac{1}{f'_n}\right) - d_n (f(y_n) - f_n).
    \label{aux}
    \end{align}
  \item \textbf{Neta $14^{th}$ order method} $(p = 14)$:
     The iterative scheme of Neta's $14^{th}$ order method \citep{N81} is given by
     \begin{align}
     &w_n = x_n - u_n, \nonumber\\
     &z_n = w_n - \frac{f(w_n)}{f'_n} \frac{f_n + b f(w_n)}{f_n + (b - 2)f(w_n)}, \nonumber\\
     &t_n = z_n - \frac{f(z_n)}{f'_n} \frac{f_n - f(w_n)}{f_n - 3f(w_n)}, \nonumber\\
     &x_{n+1} = x_n - u_n + c f_n^2 - d f_n^3 + e f_n^4,
     \label{n14}
     \end{align}
     where
     \begin{align}
     &e = \frac{\frac{\phi_t - \phi_z}{F_t - F_z} - \frac{\phi_w - \phi_z}{F_w - F_z}}{F_t - F_w}, \nonumber\\
     &d = \frac{\phi_t - \phi_z}{F_t - F_z} - e \left(F_t + F_z \right), \nonumber\\
     &c = \phi_t - d F_t - e F_t^2,
     \label{aux2}
     \end{align}
     where we use the notations
     \begin{align}
     &\delta = \delta_n - x_n, \nonumber\\
     &F_{\delta} = f(\delta_n) - f_n, \nonumber\\
     &\phi_{\delta} = \frac{\delta}{F_{\delta}^2} - \frac{1}{F_{\delta}f'_n},
     \label{aux3}
     \end{align}
     for $\delta = w, z, t$. In our computations we have used the value $b = 2$.
  \item \textbf{Neta $16^{th}$ order method} $(p = 16)$:
    The iterative scheme of Neta's $16^{th}$ order optimal method \citep{N81} is given by
    \begin{align}
    &y_n = x_n - u_n, \nonumber\\
    &z_n = y_n - \frac{f(y_n)}{f'_n} \frac{f_n + \beta f(y_n)}{f_n + (\beta - 2)f(y_n)}, \nonumber\\
    &t_n = x_n - u_n + c_n f_n^2 - d_n f_n^3, \nonumber\\
    &x_{n+1} = x_n - u_n + c f_n^2 - d f_n^3 + e f_n^4,
    \label{n16}
    \end{align}
    where $c_n$ and $d_n$ are given by Eqs. (\ref{aux}), while $c$, $d$, and $e$ are given by Eqs. (\ref{aux2}). In our computations we have used the value $\beta = 2$.
\end{itemize}

\end{document}